\documentclass[runningheads]{llncs}
\usepackage[T1]{fontenc}
\usepackage[shortcuts]{extdash}
\usepackage{graphicx}

\usepackage[usenames,dvipsnames,svgnames,table]{xcolor}
\usepackage[colorlinks=true,
            linkcolor=DimGray,
            urlcolor=RoyalBlue,
            citecolor=DimGray]{hyperref}

\usepackage[numbers, sectionbib, sort]{natbib}

\usepackage{multirow} %
 \usepackage{booktabs}  %
 \usepackage{tcolorbox} %

\newcommand{\gptfour}{\texttt{GPT-4o}}

\newcommand{\mistral}{\texttt{Mistral}}
\newcommand{\qwen}{\texttt{Qwen}}
\newcommand{\llama}{\texttt{Llama3}}
\newcommand{\gemma}{\texttt{Gemma2}}
\newcommand{\myparagraph}[1]{\vspace{0.5\baselineskip}\noindent{\textbf{#1}}.~}

\begin{document}
\title{Revisiting Human-vs-LLM judgments using the TREC Podcast Track \thanks{The paper has been accepted to appear at ECIR 2026.}}

\author{Watheq Mansour \inst{1} \orcidID{0000-0002-9463-595X} \and
J. Shane Culpepper \inst{1} \orcidID{0000-0002-9463-595X} \and
Joel Mackenzie \inst{1} \orcidID{0000-0001-7992-4633} \and
Andrew Yates \inst{2} \orcidID{0000-0002-5970-880X}}
\institute{The University of Queensland, Brisbane, Australia \and HLTCOE, Johns Hopkins University, Baltimore, Maryland, USA}
\authorrunning{W. Mansour et al.}
\maketitle

\begin{abstract}

Using large language models (LLMs) to annotate relevance is an increasingly important technique in the information retrieval community.
While some studies demonstrate that LLMs can achieve high user agreement with ground truth (human) judgments, other studies have argued for the opposite conclusion. 
To the best of our knowledge, these studies have primarily focused on classic ad-hoc text search scenarios.
In this paper, we conduct an analysis on user agreement between LLM and human experts, and explore the impact disagreement has on system rankings.
In contrast to prior studies, we focus on a collection composed of audio files that are transcribed into two-minute segments -- the TREC 2020 and 2021 podcast track.
We employ five different LLM models to re-assess all of the query-segment pairs, which were originally annotated by TREC assessors. 
Furthermore, we re-assess a small subset of pairs where LLM and TREC assessors have the highest disagreement, and found that the human experts tend to agree with LLMs more than with the TREC assessors.
Our results reinforce the previous insights of Sormunen in 2002 -- that relying on a single assessor leads to lower user agreement -- and emphasize the subjectivity of relevance assessment tasks.

\keywords{Large language models  \and Relevance judgment }
\end{abstract}

\section{Introduction}
As researchers increasingly embrace the use of large language models (LLMs), the Information Retrieval (IR) community is now vested in understanding how reliable LLMs are at labeling data~{\cite{benchllm, bmq25-sigir, gd+24-cacm, am+24-emtcir, ms23-sigir, judgeblender}}.
Much of this work has been empirical in nature, where LLM assessors judge query-document pairs that have ground truth judgments created by human assessors, allowing us to compare LLM versus human performance {\cite{faggioli2023perspectives, ts+24sigir, UMBRELA2024, upadhyay2024large}}.
These studies typically consider ad-hoc passage or document retrieval tasks such as the TREC Deep Learning track, where the MSMARCO corpus is used as the document collection~{\cite{MS_MARCO_v3}}.
We are therefore motivated to consider other related tasks, in particular, collections with different properties, to understand how both the collection and retrieval task influence LLM assessments. To this end, we explore the use of LLMs to assess relevance using the TREC podcast track, where the data is more challenging for both human and LLM judges since the collections contain transcription errors, and there is a loss of context across podcast segments. 
Our main contributions are as follows:

(1) We re-assess $18{,}284$ pairs from the Podcast Track 2020/2021 pools using five different LLMs, resulting in a total of $91{,}420$ judged pairs.
(2) We exhaustively study assessor agreement and the impact on system ordering. 
(3) We identify a set of pairs with the highest disagreement between the original TREC assessors and the LLMs, and re-assess a subset of these topics using three senior IR research experts to better understand why LLMs and TREC judges disagree.

We find that the system ordering from the 2020 collection is relatively stable when using LLM assessments, with the top 3 system orderings remaining the same; however, the 2021 system ordering changes substantially.
Surprisingly, we find that assessments from the TREC assessors systematically disagree with both the independent human experts and the LLM assessments. Initial analysis of the system descriptions suggests that LLMs favor lexical-based systems more than dense systems {\cite{alaofi2024llms}}.
These findings reveal important issues in the original 2021 assessments and illustrate how challenging assessing relevance with podcast data can be, highlighting the often ambiguous nature of relevance assessment {\cite{sormunen}}.

\section{Related Work}
We briefly outline the key work from this emerging area of research; for a more comprehensive overview, we refer the reader to the report from the first {\emph{LLM4Eval}} workshop {\cite{llm4eval}} and to recent perspective papers  {\cite{clarke2024llmcannot, soboroff2024, gd+24-cacm}}. 

In 2023, \citet{faggioli2023perspectives} explored the advantages and disadvantages of using an LLM to generate relevance judgments with an IR test collection, observing highly correlated {\emph{system orderings}} despite exhibiting only modest levels of judgment agreement between the two approaches. 
\citet{UMBRELA2024} reproduced Thomas et al's work \cite{ts+24sigir} using OpenAI's GPT-4o, and released an open-source toolkit called \emph{UMBRELA}.
Experiments using the TREC Deep Learning Tracks from 2019–2023 demonstrate that system rankings created with LLM labels are highly correlated to the ordering produced by human ground-truth judgments.
Interestingly, \citeauthor{UMBRELA2024} show several corner cases where LLM judgments were actually more accurate than the corresponding human judgments, which they attributed to the unreliability of human assessments, or a lack of a clear description of the user's information need. 

\citet{upadhyay2024large} compared LLM and human judgments using three different configurations to measure the effect of including humans during the labeling process.
The key observation is that \emph{LLM judgments could replace human assessments when using many common IR effectiveness metrics, when the overall effectiveness ordering at the system run level is being measured}. 
When comparing agreement at the judgment level, they found that human assessors apply more stringent relevance criteria than LLMs currently do -- meaning that LLMs tend to {\emph{over-rate}} relevance compared to humans.
In contrast, both \citet{clarke2024llmcannot}, and \citet{soboroff2024}, argue against the replacement of humans with LLMs, and provide counterexamples to demonstrate the pitfalls of doing so. 
The authors argue that there is no clear line between an LLM judge and an LLM re-ranker.

\section{Experimental Setup}
\vspace{-3mm}

\myparagraph{Collection and Queries}
The Spotify podcast corpus provides the ``documents'' used in the 2020 and 2021 TREC Podcast collections {\cite{clifton2020100,jones2020podcasts, Karlgren2021podcasts}}.
This dataset consists of $100{,}000$ English podcasts published between 2019 and 2020 on the Spotify platform; the episodes constitute around $60{,}000$ hours of audio. 
The audio collection was transcribed using Google's Speech-to-Text API and then partitioned into two-minute segments (each with a one-minute overlap) to form the final text collection, producing around $3.4$ million text segments (documents).
Note that this automatic transcription process differentiates the podcast corpus from ad-hoc text corpora, as it often contains errors inherent to audio transcription {\cite{mansour25ecir}}, and segments do not necessarily align with a single context like a passage-based collection does (a segment can start halfway through a sentence, for example, which can be problematic to both humans and LLMs).

The 2020 and 2021 TREC podcast tracks each contain 50 topics. 
As is typical with TREC topics, each is accompanied by a short ``title'' query, and a longer description of the user information need. The relevance judgments were generated by \emph{one assessor per topic}. NIST assessors had access to both the ASR transcript (including text before and after the text of the two-minute segment) and the corresponding audio segment.

\myparagraph{Re-Assessing the Judgment Pool}
We employ multiple open-source and proprietary LLMs 
to ensure that our findings are consistent. 
We use OpenAI’s {\gptfour} as a proprietary model, as it has achieved the highest agreement with human judgments in recent studies~\citep{alaofi2024llms, upadhyay2024large}. 
For our open-source LLMs, we use the following four models:
(1) {\mistral}, The {\tt{Mistral-Small-Instruct-2409-Q6\_K\_L}} model from the wider Mistral family {\cite{mistral}};
(2) {\qwen}, The {\tt{Qwen2.5-14B-Instruct\-/Q8\_0}} model~{\citep{qwen25technicalreport}};
(3) {\llama}, Meta's {\tt{Meta-Llama-3.1-8B-Instruct-Q8\_0}} model {\citep{llama3}}; and
(4) {\gemma}, Google's {\tt{gemma-2-9b-it-Q8\_0}} {\citep{gemma}}.
All models are quantized to $8$ bits with the exception of {\mistral} which is a $6$-bit model.
We use 
llama-cpp-python\footnote{\url{https://github.com/abetlen/llama-cpp-python}} to support more efficient inference at scale. 
All of the open source models are publicly available.\footnote{\url{https://huggingface.co/collections/bartowski/}}

\myparagraph{LLM Prompting} Before running the LLMs on the assessment pool, we fine-tune our instruction prompt using a 10\% stratified random sample of TREC query-segment pairs.
In our initial prompt, we included the TREC judgment guidelines in the \emph{Description, Narrative, Aspects} (DNA) prompting style as it achieves the best performance on relevance assessment tasks~\citep{ts+24sigir}. 
The descriptions and narratives assist LLM in understanding the topic intent, and the retrieved segment that should be assessed, whereas the aspects guide the thinking process in a step-by-step manner. 
The output is restricted to a JSON object that contains the relevance score and a short justification {\cite{tam2024let}.
We experiment using three variants of the original prompt: a vanilla zero-shot DNA prompt; a prompt that asks LLMs to be more strict in its assessments, inspired by \citet{upadhyay2024large} who show that LLMs tend to overestimate relevance compared to humans; and a prompt using in-context learning.
We evaluated the quality of the prompts using Cohen Kappa's score relative to the ground-truth judgments from TREC, and then chose the prompt that had the highest agreement. %
We used the best-performing prompt (the second variant) in all subsequent experiments.

\myparagraph{Normalizing Relevance Grades} 
According to the TREC judgment guidelines, there are five relevance grades (0-4), and grade 4 was used only for ``known item'' and ``refinding'' topic types (and not for the ``topical'' category).
However, upon examining the judgment pairs from both 2020 and 2021, we found that grade 4 was applied on every topic type, making it unclear how to differentiate between grades 3 and 4 based on the categorical description of each. 
In addition, perfect relevance (grade 4) is not reproducible by anyone but the topic creator. 
Therefore, we remap all such pairs with a grade of 4 to 3 to provide a more stable testing framework.
Thus, we use a four-point relevance scale, which aligns with the graded relevance range used in previous TREC ranking tasks, such as the Deep Learning passage and document ranking tasks~\citep{trecdl21, trec2020, trec2019deeplearning}.

\myparagraph{Reproducibility}
In the interest of reproducibility, we make the used prompts and LLMs' judgments publicly available. \\[2ex]
\vspace{1em}
\hspace{3.8em}\includegraphics[width=1.25em,height=1.25em]{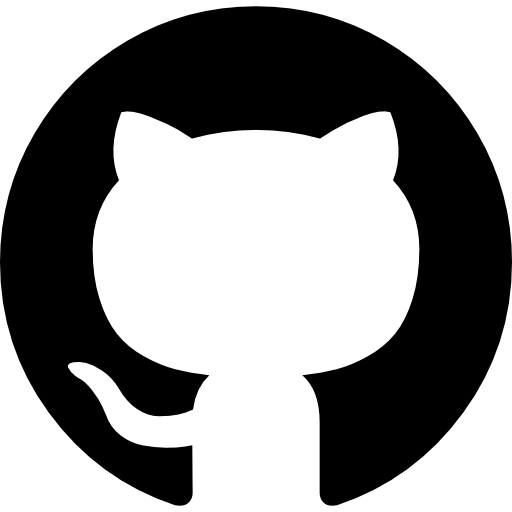}\hspace{.5em}
\parbox[c]{\columnwidth}
{
    \vspace{-.55em}
    \textcolor{RoyalBlue}{\nolinkurl{https://github.com/Watheq9/ecir2026-podcast-judge}}
}

\section{Results}
\vspace{-2mm}

After exhaustively reassessing the entire set of query-document pairs derived from the TREC 2020 and 2021 podcast tracks using five LLMs, we evaluate how these judgments compare to the ground truth judgments from TREC.
In the experiments presented below, all systems are ordered according to the mean RBP $\phi = 0.95$ score {\cite{rbp}}; the same trends were observed using NDCG@10 {\cite{ndcg}}.

\begin{table}[t]
\setlength{\tabcolsep}{0.7em}
\caption{Run order correlations for the TREC 2020 pool (28 systems, left) and the TREC 2021 pool (27 systems, right) when comparing human assessments to various LLMs.}
\label{tab-kendall-rbp}
\centering
\begin{tabular}{l ccc ccc}
\toprule
&  \multicolumn{3}{c}{TREC 2020} & \multicolumn{3}{c}{TREC 2021} \\
\cmidrule(lr){2-4}\cmidrule(lr){5-7}
\multirow{2}{*}{Model}  & \multirow{2}{*}{Kendall's $\tau$} & \multicolumn{2}{c}{RBA $\phi$} & \multirow{2}{*}{Kendall's $\tau$} & \multicolumn{2}{c}{RBA $\phi$} \\
\cmidrule(lr){3-4}\cmidrule(lr){6-7}
&  & $0.8$ & $0.9$  & & $0.8$ & $0.9$ \\
\midrule

GPT-4o & 0.81 & 0.96 & 0.94 & 0.66 & 0.94 & 0.92\\
Mistral & 0.84 & 0.97 & 0.94 & 0.67 & 0.94 & 0.92\\
Qwen & 0.83 & 0.97 & 0.94 & 0.62 & 0.94 & 0.92\\
Llama3 & 0.81 & 0.96 & 0.94 & 0.72 & 0.95 & 0.93\\
Gemma2 & 0.84 & 0.97 & 0.94 & 0.60 & 0.92 & 0.91 \\
\bottomrule
\end{tabular}
\end{table}

\myparagraph{System Ranking Evaluation}
Table~\ref{tab-kendall-rbp} reports the system ranking correlations using both an unweighted Kendall's $\tau$ {\cite{kendall}}, and the top-weighted {\emph{Rank-Biased Alignment}} (RBA) {\cite{rba}}. 
Two different settings for RBA are used: (1) a shallow version ($\phi = 0.8$, representing an expected depth of $5$), focusing the weight of the comparison at the top of the ranking; and (2) a deeper version ($\phi = 0.9$, representing an expected depth of $10$), spreading the weight more uniformly across all system rankings.

In the 2020 comparison, the results are quite stable, with high agreement between the system orderings produced using human relevance assessments compared to the LLM assessments, regardless of the metric or LLM being applied. 
In particular, observe that Kendall's $\tau$ values are as high as $0.84$, and that the top-weighted RBA metric always returns results greater than $0.94$, indicating that the top ranking system ordering is being preserved between the human and LLM judges. 
However, the 2021 judgments present a much different story, with Kendall's $\tau$ values as low as $0.60$, and lower RBA values in all of our comparisons. 
This indicates that the system ordering is much more volatile than in the 2020 data, including the {\emph{top}} ranking systems.

To illustrate how the system ordering volatility as the LLM assessor is changed, we plot the changes in rank position for each system, compared to the human judgments -- Figure~\ref{fig-rbp-rank} shows the results for both 2020 (left) and 2021 (right), which align with Table~\ref{tab-kendall-rbp}. 
In 2020, it is clear that the top ranking system ordering is largely preserved, with small perturbations occurring after the third-best system. 
However, the systems' rank changes are remarkably larger in 2021. The third- and fourth-ranked systems drop between one and five positions depending on the LLM used to create the judgments, the fifth-best system moves two positions up, and the eighth system drops up to {\emph{twelve}} positions. 
Even more surprising is that the system that is originally ranked at position 13 moves into the top six systems, with similar large positive deltas observed as deep as rank 19.
Initial analysis of the runs in Figure~\ref{fig-rbp-rank} suggests that LLM judgments may favor lexical systems (including hybrids or re-rankers with lexical first stages) as compared to strictly dense systems {\cite{alaofi2024llms}}.
For example, the \emph{QL} baseline jumped from rank $18$ to $16$, and from $22$ to $15$ in 2020 and 2021, respectively (See Gemma2 in Figure~\ref{fig-rbp-rank}). However, more research is required to 
completely understand the instability of the 2021 data.

\begin{figure*}[t!]
\centering
\includegraphics[width=0.5\textwidth]{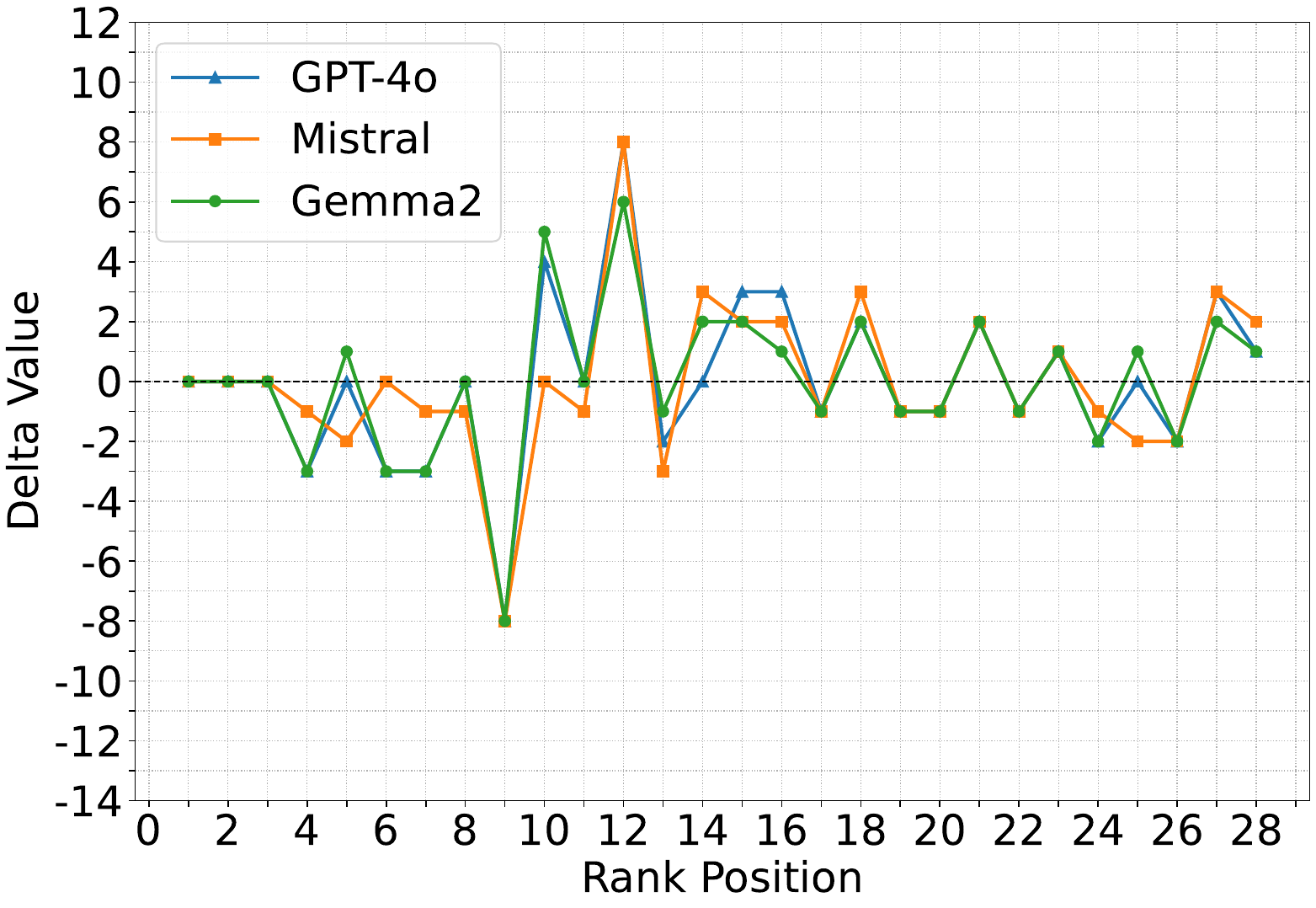}\includegraphics[width=0.5\textwidth]{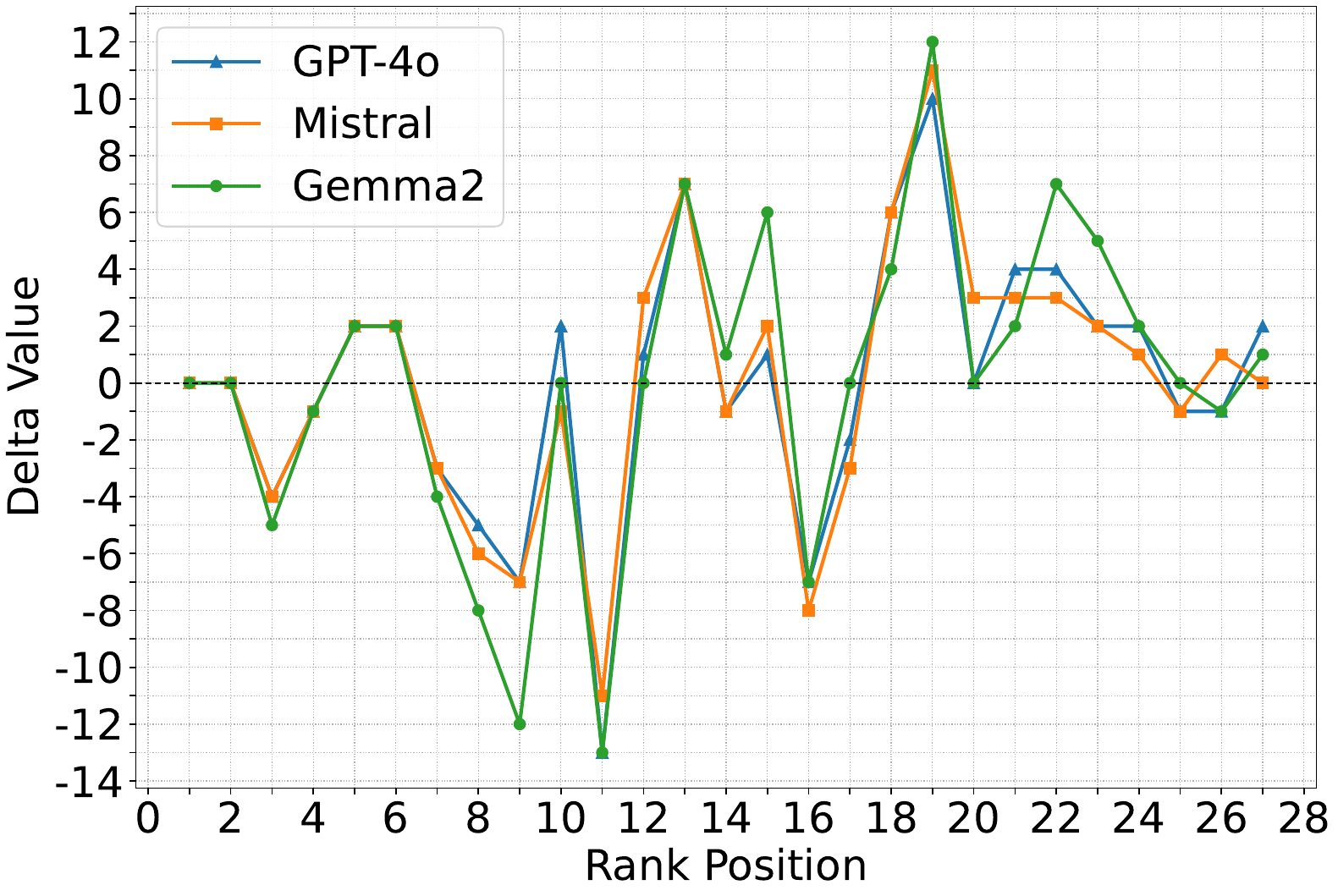}
\caption{Order volatility in TREC systems scored using RBP $\phi = 0.95$ in 2020 (left) and 2021 (right), when ordering changes using LLM assessments. The $y$-axis expresses the difference in rank position compared to the ground truth system ordering.}
\label{fig-rbp-rank}
\end{figure*}

\myparagraph{Human Assessor Agreement}
To better understand when the LLM assessors disagree with the TREC assessors, we randomly sampled $22$ out of $826$ query-document pairs representing ``high disagreements'' -- where the absolute difference in label between the TREC assessors and the (majority vote) LLM assessors was greater than two. 
Then, three IR experts independently judged these pairs. 
The annotation guidelines were the same prompt used when judging with LLMs, with additional information about query types. The pairs were provided in a sheet, where each pair was represented by the query, topic description, and a transcript of a 2-minute audio segment.
Table~\ref{tab-kripp-subset} shows the inter-rater agreement between each expert assessor, the TREC assessor, and the LLM assessments.
Surprisingly, the agreement between the expert assessors and the LLMs falls in the {\emph{tentative}} to {\emph{reliable}} range; on the other hand, there is a systematic disagreement between the TREC assessors and both the human assessors and the LLMs.
This supports the ealier findings of \citet{sormunen} who also demonstrated that the ambiguity of relevance assessments can result in vastly different outcomes for a query-document pair -- and, in this context, it suggests that (many) LLMs may be more reliable than (one) human {\cite{UMBRELA2024}} and are clearly more reliable than other related work suggests.
In total, $234$ out of $9{,}386$ pairs in 2020 (2.5\%), and $1{,}014$ out of $8{,}897$ pairs in 2021 (11.4\%) had TREC assessor assign a $0$ label, compared to all of the LLM labels that were  $\geq 2$. 
The converse (when the TREC label is a $2$ or $3$ and the LLM label is a $0$) occurs in a much smaller number of disagreements -- $11$ and $28$ pairs in 2020 and 2021, respectively -- corroborating the notion that LLMs tend to assign higher relevance to a pair than humans {\cite{upadhyay2024large}}, and providing a potential explanation for at least some of the instability observed for the 2021 collection.

\begin{table}[t]
\setlength{\tabcolsep}{0.7em}

\caption{Inter-rater agreement as measured with Krippendorff’s $\alpha$
between assessments of three senior human annotators, LLMs (majority vote), and the TREC judgments across the sampled query-document pairs from the TREC pool.}
\label{tab-kripp-subset}
\centering
\begin{tabular}{r r r r r}
\toprule
  & Annotator 2 &  Annotator 3 & TREC & LLMs \\
\midrule
Annotator 1 & 0.67 & 0.73 & $-0.66$ & 0.71 \\
Annotator 2 & -- & 0.82 & $-0.77$ & 0.86 \\
Annotator 3 & -- & -- & $-0.55$ & 0.77 \\
TREC & -- & -- & -- & $-0.76$ \\
\bottomrule
\end{tabular}
\end{table}

\section{Conclusion}
We have revisited using LLMs as relevance assessors.
We found that, although the correlation between the TREC and the LLM assessors was high in the 2020 TREC Podcast collection, it was much more volatile in the 2021 collection, raising doubts about the stability of the gold label assessments. 
Our analysis indicates that the LLM assessments tend to favor lexical systems, causing them to score much higher in system ranking comparisons.
We also had three IR experts independently reassess a subset of pairs where the TREC and LLM judgments had the highest disagreement, and found that the new human judgments have a much higher agreement with the LLM labels than with the original human judgments. 
This preliminary work corroborates a number of recent findings on LLMs for relevance assessments using two new test collections, and further emphasizes the ambiguous nature of relevance assessment tasks. 
We plan to continue our analysis to better understand the instability we observed on the 2021 TREC Podcast campaign in future work.

\begin{credits}
\subsubsection{\ackname}
 We thank the anonymous referees for their feedback and suggestions. The third author was supported by a Google Research Scholar grant.

\subsubsection{\discintname}
The authors have no competing interests of any sort.
\end{credits}

\bibliographystyle{splncs04nat}
\bibliography{references}

\end{document}